\documentclass[referee]{cjaa}           

\usepackage{graphicx}                   
\input{epsf.sty}                        
\input{psfig.sty}                       

\begin{document}

   \title{News from Galactic Black Holes
}

\ifx \volnopage \undefinedmacro
    \def \volnopage#1{}
\fi

   \volnopage{Vol.0 (200x) No.0, 000--000}      
   \setcounter{page}{1}          

   \author{Janusz Zi\'o{\l}kowski
      }
   \offprints{J. Zi\'o{\l}kowski}                   

   \institute{Copernicus Astronomical Center, ul. Bartycka 18, 00-716 Warsaw, Poland\\
             \email{jz@camk.edu.pl}
          }

   \date{Received~~2003 month day; accepted~~2003~~month day}

   \abstract{
Present status of the possible black hole microlensing events and
the present understanding of high frequency quasi-periodic
oscillations in systems containing black hole candidates (BHCs)
are briefly discussed. The growing evidence for the presence of
the event horizons around some compact objects is reviewed.
Finally, the news from four individual objects (SS 433, GX
339$-$4, CI Cam and Cyg X-3) are presented and the updated list of
BHCs, containing 49 objects is given.
   }

   \authorrunning{J. Zi\'o{\l}kowski}            
   \titlerunning{News from Galactic Black Holes}  

   \maketitle
%
%
\section{Introduction}           
\label{sect:intro}

In agreement with the title of my talk, I will briefly consider
the new data and/or new interpretations in different fields of
black hole research. The selection of the topics will be,
necessarily, arbitrary. First, I will mention the actual status of
black hole candidates (BHCs) from microlensing events. Next, I
will discuss the present situation in the field of the
investigation of high frequency quasi-periodic oscillations (QPOs)
observed in galactic BHCs. Then, I will review the observational
evidence indicating the presence of the event horizons around
compact objects in some X-Ray Novae (XNe). Finally, I will discuss
four individual systems containing confirmed or possible BHCs: SS
433, GX 339$-$4, CI Cam and Cyg X-3. At the end, I will present
the updated table of galactic BHCs. At present, the table contains
49 objects, 19 of which may be considered confirmed black holes.


\section{Black Hole Candidates from Microlensing Events}
\label{sect:mlev}

Among several hundreds of microlensing events observed so far
there are few long events with the time scales of months to years.
The long time scale of a microlensing event may be attributed
either to a large mass of the lens or to the small relative
transverse velocity of the lens with respect to the lensed object
and the observer. The long events show, usually, the magnification
fluctuations, reflecting the motion of the Earth. This effect
permits to calculate the so called "microlensing parallax" which
is a measure of the relative transverse motion of the lens with
respect to the observer. Assuming standard model of the Galactic
velocity distribution, one is then able to perform a likelihood
analysis, which permits to estimate the distance and the mass of
the lens. With the help of the above analysis, three long events
were selected as, possibly, caused by black hole lenses. The
candidates are: MACHO-98-BLG-6 (probable mass of the lens $\sim 3
\div 13 M_\odot$, Bennett et al., 2002a), MACHO-96-BLG-5 (probable
mass of the lens $\sim 3 \div 16 M_\odot$, Bennett et al., 2002a)
and MACHO-99-BLG-22 = OGLE-1999-BUL-32 (probable mass of the lens
$\sim 100 M_\odot$, Bennett et al., 2002b). Only the last of them
seems to be a robust candidate. I will also add, that Paczy\'nski
(2003) promises more BH lenses from OGLE project in some 2 $\div$
3 years.

\section{High Frequency QPO's  in Black Hole Candidates}
\label{sect:qpo}

The list of high frequency QPOs observed in X-ray emission from
galactic BHCs is growing fast in recent years. The actual list is
given in Table 1 (compiled from Remillard et al., 2002 and
McClintock and Remillard, 2003). The most striking feature of
these QPOs is the fact, that in most of the systems the QPO
frequencies form sets of  precise integral harmonics. The
fundamental frequency seems to be unique characteristic of each
black hole and, presumably, depends only on its mass and spin. We
observe 2:3 harmonics in GRO J1655$-$40 and XTE J1550-564 and
1:2:3 harmonics in GRS 1915+105. This last system shows,
additionally, an independent set of 3:5 harmonics (41 and 67 Hz).
One can also speculate about set of 113 and 164 Hz QPOs (2:3).
There are different proposed mechanisms to explain the origin of
high frequency QPOs observed in BHCs (one should add that BHCs
exhibit also low frequency QPOs with frequencies in the range 0.1
$\div$ 10 Hz, which must be of different origin). At present, the
leading theories are: the parametric epicyclical resonance in the
inner accretion disc (Abramowicz and Klu\'zniak, 2001) or the
different modes of oscillations of the inner disc (so called,
discoseismology $-$ see e.g Wagoner et al., 2001). Both theories
predict that the frequencies should scale with the mass of the
compact object like $M^{-1}$. Looking at Table 1, one can check
that QPOs in GRO J1655$-$40, XTE J1550-564 and XTE J1859+226 have
the proper scaling but GRS 1915+105 is, rather, an exception
(unless one selects 113 Hz QPO as an equivalent of 300, 184 and
193 Hz in the other three sources). In addition, GRS 1915+105 has
at least two sets of high frequency QPOs and these two sets,
probably cannot be explained by a single mechanism. The integral
harmonics are predicted, in a natural way, only by the epicyclical
resonance mechanism and, in this respect, the observations,
clearly, support this theory. On the other hand, one should
remember that the oscillating (epicycling) element, emitting
X-rays, experiences substantial damping forces. There are serious
doubts, whether the element, following the epicyclical motion, can
preserve its identity long enough to produce the observable signal
seen as QPOs (see, e.g. Markovic and Lamb, 1998, 2000). Both the
epicyclical resonance mechanism and the discoseismology predict
that the frequencies, in addition to scaling with the mass, should
grow with the increasing spin of the black hole. Therefore,
regardless of which explanation will appear finally to be the
correct one, the high frequency QPOs will remain an important
diagnostic tool in estimating the masses and the spins of the
accreting BHs and, also, in distinguishing between accreting BHs
and accreting NSs.

\begin{table}[]
  \caption[]{High Frequency QPOs observed in BHC Binary Systems.}
  \label{Tab:qpot}
  \begin{center}\begin{tabular}{|rcl|rcl|rcl|c|}
\hline
&&&&&& &&&\\
\multicolumn{3}{|c|}{Name}&\multicolumn{3}{|c|}{$\nu_{qpo}$ }&\multicolumn{3}{|c|}{$M_{BH}$}&\multicolumn{1}{|c|}{comments}\\
&&&\multicolumn{3}{|c|}{[Hz]}&\multicolumn{3}{|c|}{[M$_\odot$]}&\\
\hline
&&&&&&&&&\\
GRO J1655\hspace*{-3ex}&$-$&\hspace*{-3ex}40&300\hspace*{-2.4ex}&$\pm$&\hspace*{-2.4ex}23&6\hspace*{-3ex}&.&\hspace*{-3ex}3&\\
&&&450\hspace*{-2.4ex}&$\pm$&\hspace*{-2.4ex}20&&&&\\
XTE J1550\hspace*{-3ex}&$-$&\hspace*{-3ex}564&184\hspace*{-2.4ex}&$\pm$&\hspace*{-2.4ex}26&10\hspace*{-3ex}&.&\hspace*{-3ex}6&\\
&&&272\hspace*{-2.4ex}&$\pm$&\hspace*{-2.4ex}20&&&&\\
GRS 1915\hspace*{-3ex}&+&\hspace*{-3ex}105&41\hspace*{-2.4ex}&$\pm$&\hspace*{-2.4ex}1&14\hspace*{-3ex}&&&\\
&&&67\hspace*{-2.4ex}&$\pm$&\hspace*{-2.4ex}5&&&&\\
&&&113\hspace*{-2.4ex}&&&&&&\\
&&&164\hspace*{-2.4ex}&$\pm$&\hspace*{-2.4ex}2&&&&\\
&&&328\hspace*{-2.4ex}&$\pm$&\hspace*{-2.4ex}4&&&&\\
&&&495\hspace*{-2.4ex}&&&&&&1.5 $\sigma$\\
4U 1630\hspace*{-3ex}&$-$&\hspace*{-3ex}472&184\hspace*{-2.4ex}&$\pm$&\hspace*{-2.4ex}5&&&&\\
XTE J1859\hspace*{-3ex}&+&\hspace*{-3ex}226&193\hspace*{-2.4ex}&$\pm$&\hspace*{-2.4ex}4&9\hspace*{-3ex}&&&\\
H 1743\hspace*{-3ex}&$-$&\hspace*{-3ex}322&240\hspace*{-2.4ex}&&&&&&\\
XTE J1650\hspace*{-3ex}&$-$&\hspace*{-3ex}500&250\hspace*{-2.4ex}&&&&&&\\
&&&&&&&&&\\
\hline
  \end{tabular}\end{center}
\vspace{10mm}
{\small NOTE: 495 Hz QPO in GRS 1915+105 was detected at only 1.5 $\sigma$ significance level}

\end{table}

\section{Evidence for the Presence of the Event Horizons around Compact Objects in some X-Ray Novae}
\label{sect:evhor}

In this section, I will review the, growing recently, evidence indicating that compact components in some X-ray Novae most likely posses event horizons around them.

\subsection{Missing Type I Bursts}

X-ray Novae (XNe, called also Soft X-Ray Transients or SXTs)
change their accretion rates (as inferred from the X-ray
luminosities) by many orders of magnitude. Therefore, there must
be time intervals, when the accretion rates fall into certain
critical range of values, corresponding to the presence of the
Type I bursts (thermonuclear flashes due to unstable nuclear
burning of the accreted matter on the surface of the NS). About 50
XNe are known at present. In roughly 85 \% of them, the compact
object is a BHC, in the remaining $\sim$ 15 \% it is a NS. In most
of the NS XNe, Type I bursts are observed from time to time.
However, not a single Type I burst was ever seen in any of the BHC
XNe. This is a strong argument in favor of the presence of the
event horizon, since Type I bursts can occur only if the solid
surface is present. Narayan and Heyl (2002) investigated the
stability of the nuclear burning on the surfaces of accreting
compact objects of 1.4 and 10 $M_\odot$ for different accretion
rates (assuming the presence of the solid surface in each case).
In agreement with many earlier investigations, they found
instability for some accretion rates in the case of 1.4 $M_\odot$
object (corresponding to the typical NS). But they found
instability also for the 10 $M_\odot$ object. It seems, therefore,
that independently of the mass, if solid surface is present, Type
I bursts must appear. More recently, Narayan and Heyl (2003) made
additional modeling, investigating the recurrence time of Type I
bursts from a hypothetical 10 $M_\odot$ compact object, possessing
a solid surface. They found quite short recurrence times ($\sim
0.5 \div 7$ days), similar to those of typical burst from NSs.
These bursts occur at somewhat higher luminocities ($\sim 0.1 \div
0.6$ Eddington luminocities vs $\sim 0.02 \div 0.2$ for NSs), but
the luminocities in question are observed for the erupting XNe.
Also the calculated durations of the bursts (100 to 500 s vs 20 to
300 s for NSs) should permit easy observation of these bursts. The
fact that not a single burst was ever seen implies, therefore, the
presence of the event horizon around the compact object.

\subsection{Some Quiescent X-Ray Novae are "Blacker" than some Others}

Between the outbursts, XNe retain some (although very low) level
of X-ray luminosity, which indicates that some marginal accretion
is still present. This, so called "quiescent luminosity" is so
low, that it was not detectable, until the recent generation of
the X-ray satellites. The comparison of the quiescent luminosities
of BHC XNe and NS XNe shows a very substantial difference between
the two classes of objects. For comparable orbital periods (which
implies comparable sizes of the orbits and, so, the comparable
accretion rates), the BHC XNe are, systematically, by about two
and a half $-$ three orders of magnitude dimmer than NS XNe
(Narayan  et al. 2001, Garcia et al. 2001, Narayan 2003). The most
obvious explanation of this difference is the presence of the
solid surface (where most of the accretion energy is released) in
the NS XNe. BHC XNe are "blacker" because they have event
horizons. This relative dimness of BHC XNe is well seen in Fig. 1
(adapted from Narayan, 2003).

There are some alternative interpretations of the dimness of the
quiescent BHC XNe: the consumption of the accretion power by the
jets (Fender et al., 2003) or the "gravastars" (Mazur et Mottola,
2001), but they are far less convincing than the most conservative
interpretation: the presence of the event horizons.

\begin{figure}
   \vspace{2mm}
   \begin{center}
   \hspace{3mm}\psfig{figure=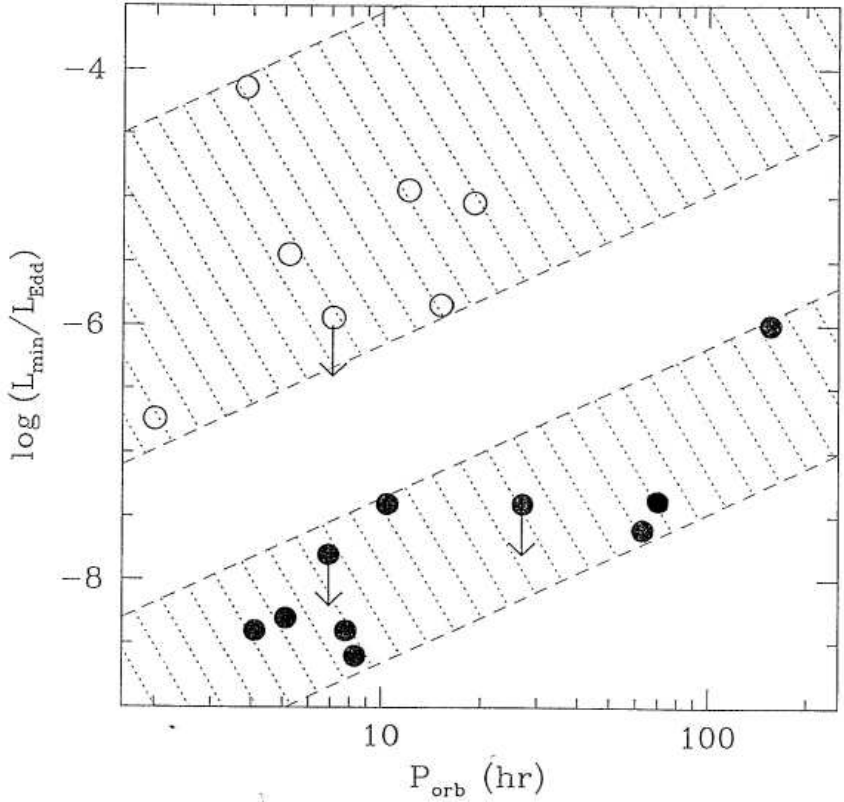,width=120mm}
   \caption{Quiescent luminosities (in Eddington units) plotted against orbital period for 17 XNe. The open circles correspond to systems containing NSs and  the filled circles to systems containing BHCs. The picture is an updated version of Fig.2 from Narayan (2003). Data for the point corresponding to the system XTE J1819$-$254/V4641 Sgr are from McSwain et al. (2003).}
   \label{Fig:quiesc.lum}
   \end{center}
\end{figure}

\subsection{Separation in the X-Ray Colour$-$Colour Diagram}

Done and Gierli\'nski (2003) analyzed huge archive of RXTE
observations to investigate the evolution (with the varying
accretion rate) of different X-ray sources in the X-ray
colour$-$colour diagram. This investigation was done for all
binary X-ray sources (not only for XNe). One of their most
striking results was the fact that positions of BHCs and NSs are
quite well separated. This is well illustrated in Fig. 2. While
there exist some small regions of overlapping, most of the BHCs
positions (in soft states) occupy a region of the diagram (the
hatched area in Fig. 2) which is never visited by any neutron
star. It is obvious, that in soft states, the spectra of BHCs are
substantially softer than the spectra of NSs. The most
conservative explanation is again that NSs have solid surfaces and
boundary layers where most of the accretion energy is released and
this makes their spectra harder, while BHCs have event horizons
and, therefore, softer spectra.

\begin{figure}
   \vspace{2mm}
   \begin{center}
   \hspace{3mm}\psfig{figure=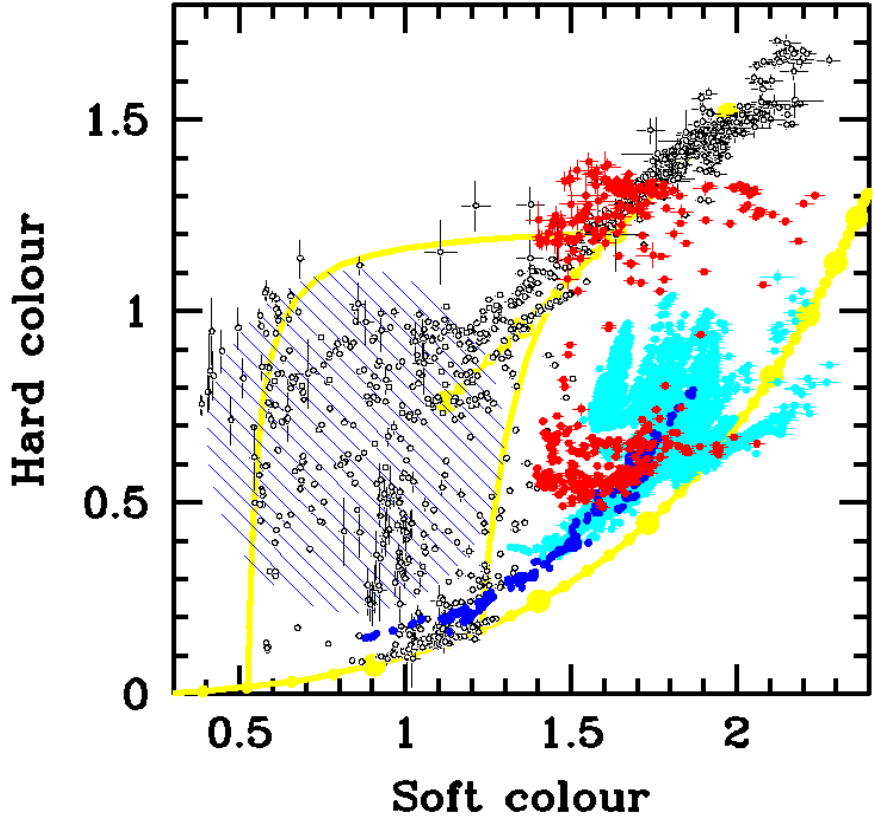,width=120mm}
   \caption{Colour-colour diagram for RXTE observations of different X-ray soures. Open circles correspond to sources containing BHCs, red, cyan and blue filled circles to different types of sources containg NSs. The hatched region is accessible only for BHCs. The picture is reproduced from Done and Gierli\'nski (2003)}
   \label{Fig:c-cdiagram}
   \end{center}
\end{figure}

\subsection{Disappearing Inner Disc}

Multicolor blackbody fits made to the disc component of the X-ray spectrum, as the spectrum evolves due to varying accretion rate, seem to indicate that in some sources the inner part of the disc suddenly disappears (the inner radius of the disc, calculated from the fit, suddenly increases by a factor of few). In most cases, this happens at low accretion rates and therefore the disappearance might be only apparent: the accretion flow in the inner disc might be just replaced by another type of flow with low radiative efficiency (e.g. ADAF). However, this cannot be the case with GRS 1915+105, where the accretion rate is always at least half of the Eddington rate or more. In the case of GRS 1915+105, the disappearance of the inner disc must be real: from time to time the substantial amount of matter is suddenly dumped on the compact object. However, we do not see the accompanying flash which should be produced as this matter hits the solid surface. This implies that there is no solid surface, but an event horizon instead.

\section{News from Individual Objects}
\label{sect:indo}

In this section, I will consider four individual BHCs, for which we obtained either new data or new ideas, which might be important for clarifying the status of the objects. For three of these objects, the BHC status was strengthened, for one (Cyg X-3) the status (BH or NS) remains vague.

\subsection{SS 433 (spectrum of the donor star seen at last!)}

This exceptional object has been celebrated for many reasons
(among others for being the first discovered galactic microquasar)
and studied very extensively. In spite of that, until recently,
there was no clear proof that the compact component of the system
is a black hole. The main reason for the troubles was the fact
that the spectrum of the companion of the compact object (the mass
donor or the "optical component") was not visible. The reason of
this "invisibility" was, in turn, the fact that the optical
emission is dominated by the accretion disc. The system is very
luminous in the optical band: $L_{\rm opt} \geq 2 \times 10^{38}$
erg s$^{-1}$. About 80 \% of this luminosity comes from the
photosphere of geometrically thick accretion disc, which can be
treated as a flattened star. The observed He II emission lines
also originate on the surface of the disc and reflect its orbital
motion. Until recently, no spectral lines produced by the second
member of the binary (the mass donor) were seen. Therefore,
atypically, only the mass function $f(M_{\rm opt})$ (where $M_{\rm
opt}$ is the mass of the unseen donor) could be estimated.
Unfortunately, the value of $f(M_{\rm opt})$ is not established
very precisely. Looking through the literature, one can find
values from 10.1 M$_\odot $ (Crampton and Hutchings, 1981) all the
way down to 1.90 $M_\odot$ (D'Odorico et al., 1991). The most
convincing, in my opinion, analysis of the emission lines from the
disc was performed by Fabrika and Bychkova (1990), who measured
the semiamplitude of radial velocities as $K_{\rm X} = 175 \pm 20$
km s$^{-1}$. This corresponds to the mass function $f(M_{\rm opt})
= 7.7 (+ 3.0, - 2.4) M_\odot$. The inclination of the orbital
plane is known very precisely from the kinematic model of the
jets: $i = 78.82^{\rm o} \pm 0.11^{\rm o}$. Unfortunately, until
recently, masses of the components could not be calculated since
the second mass function was missing (no radial velocities of the
mass donor could be measured because no lines from this component
were visible). Only circumstantial procedures could be applied.
The most important ones were the modellings of the optical light
curves (Leibowitz, 1984, Antokhina and Cherepashchuk, 1985), which
were used to constrain the mass ratio $q = M_{\rm x}/ M_{\rm opt}$
and so to constrain also the masses. The results of these modeling
were not very precise and led only to the conclusion that the mass
of the compact component $M_{\rm x}$ must be in the range of $\sim
4 \div 60 M_\odot$). This indicated the presence of a black hole
in the system, but this estimate was frequently questioned as too
uncertain.

The good news from SS 433 is that the spectrum of the second
component was finally seen! Gies et al. (2002) detected the
absorption lines of the mass donor and were able to measure their
radial velocities. They estimated the spectrum as A-type
supergiant ($\sim$ A7 Ib) and measured the semiamplitude of the
radial velocities: $K_{\rm opt} = 100 \pm 15$ km/sec (which
corresponds to $f(M_{\rm x}) = 1.36 (+ 0.71, - 0.52) M_\odot $).
Taking the semiamplitude of the radial velocities of the emission
lines from Fabrika and Bychkova ($K_{\rm X} = 175 \pm 20$ km
s$^{-1}$) and using the known value of the inclination ($i =
78.8^{\rm o} \pm 0.1^{\rm o}$), they estimated the mass ratio
$M_{\rm x}/M_{\rm opt} = 0.57 \pm 0.11$ and the masses $M_{\rm
opt} = 19 \pm 7 M_\odot$ and $M_{\rm x} = 11 \pm 5 M_\odot$.

The bad news is that the above estimate of the semiamplitude of
the radial velocities is not very reliable. The measurements cover
only small part of the orbital cycle and they were not confirmed
by the subsequent observations (Charles et al., 2004). Similarly
as in the case of Cyg X$-$3 (see section 5.4), it is difficult to
separate the effects of the orbital motion from those of the
stellar wind. While the case for the black hole presence is quite
strong, we have still to wait for the reliable estimate of its
mass.

\subsection{GX 339$-$4 (dynamical evidence for a black hole)}

This system was one of the earliest candidates for harboring a
black hole because of the similarity of the spectral and temporal
properties of its X-ray emission to those of Cyg X-1. Quite soon,
it was realized that, unlike Cyg X-1, it cannot be a massive X-ray
binary. GX 339$-$4 was occasionally entering X-ray "off" states
(never observed in Cyg X-1), during which its optical luminosity
was decreasing dramatically, indicating a low mass optical
companion. The attempts to measure radial velocities during the
"off" states were not successful because of the dimness of the
optical companion ($V \ga 21$). Only recently, its spectral type
was determined from IR observations (Chaty et al., 2002) as
F8$-$G2 III. Finally, the measurements of the probable orbital
motions of this elusive star were announced this year (Hynes et
al., 2003). Atypically, it was done not during the "off" state but
during the outburst. The lines measured were sharp NIII emission
lines believed to be formed on the surface of the optical star due
to its irradiation. These lines were found to move with the
velocities in excess of 300 km/sec. The most likely value of the
orbital period consistent with the observations is 1$^{\rm d}$756
and the estimated mass function is $f(M_{\rm x}) = 5.8 \pm$ 0.5
M$_\odot$. The broad HeII emission lines, originating probably in
the accretion disc (and reflecting the orbital motions of the
compact component) show much smaller radial velocities. The
implied mass ratio $q$ = $M_{\rm opt}/M_{\rm x}$ is $< 0.08$. Both
this low mass ratio and the implied mass of the black hole ($\ga$
6 M$_\odot$) are typical for the low mass BHC binaries.

\subsection{CI Cam (black hole or white dwarf?)}

 XTE J0421+560 (= CI Cam) was discovered as a very fast X-ray transient in 1998. Its rise time scale was only $\sim$ 12 hours and decay time scale only $\sim$ 2 weeks. The transient was classified as a microquasar, since the outburst was accompanied by the ejection of a relativistic corkscrew jets seen on the radio map. The observed motion of the jets was $\sim$ 26 mas/day, which (in connection with the present knowledge of the distance to the system) corresponds to an apparent velocity $>$ 0.75 c. The optical counterpart was identified as a long known variable symbiotic star CI Cam. Its present spectral classification is sg B[e], which implies a massive high-luminosity star undergoing a heavy mass loss through a two component wind. The source was observed in quiescent state (half a year after the outburst) by BeppoSAX. Most of the quiescent X-ray emission was found to be very soft (k$T \sim 0.22$ keV), which led the observers (Orlandini et al., 2000) to speculate that the compact object might be a white dwarf and the outburst could be due to thermal runaway on its surface (similar to classical Nova). After that paper, the system was removed from the list of BHCs. It was reinstalled again, after the paper of Robinson et al. (2002), who published the analysis of the spectroscopic observations of CI Cam. The authors observed the star two weeks after the peak of the outburst, using 2.7 m telescope at McDonald Observatory and two years after the outburst with the help of the HST. They concluded that the distance to the system must be much larger than previously estimated and is, most likely, larger than 5 kpc. It implies that CI Cam is a very bright supergiant ($M_{\rm bol} <$ -9.2). It also implies that XTE J0421+560 was a very bright X-ray transient: its peak luminocity (in 2$\div$25 keV band) had to be larger than $\sim 3 \times 10^{38}$ erg s$^{-1}$. This new distance determination leads also to the new estimates of the quiescent emission levels. The quiescent X-ray emission observed by BeppoSAX may be separated into two components: the hard one with the luminosity (in the 2$\div$10 keV band) $>$ 5 $\times 10^{32}$ erg s$^{-1}$ and a very soft (k$T \sim$ 0.22 keV) thermal component with the luminosity $>$ 2 $\times 10^{34}$ erg s$^{-1}$. The soft component is very similar to the X-ray emission from so called "supersoft sources" which are, most likely, accreting white dwarfs (see e.g. Kahabka, 2002). This similarity led Orlandini et al. to speculate about possible white dwarf nature of the compact companion of CI Cam. However, Robinson et al. point out that the hard component (the luminosity of which is $\la 1.7 \times 10^{-6}$ of the peak X-ray luminosity) is typical for the quiescent emission of the BH XNe. They argue, that most of the quiescent X-ray emission (and, especially, the entire soft component) are associated rather with the sg B[e] star and its wind, and not with the compact object. The hard component (or most of it) is probably due to remnant accretion on the compact object. This component fits very well the typical quiescent emission of the BH XNe (which suggests that the compact object is probably a black hole). The quiescent emission from XTE J0421+560 was observed again in 2002, with the help of XMM Newton (Boirin et al., 2002). The conclusions were similar to those of Robinson et al.: the unusual appearance of the quiescent emission is due to the presence of the dense circumstellar wind and most of this emission is due to sg B[e] star and its wind.

To summarize: XTE J0421+560/CI Cam binary system remains as a reasonably strong black hole candidate in the class of XNe.

\subsection{Cyg X-3 (black hole or neutron star?)}

Cyg X-3 is one of the most investigated X-ray binaries, yet we are still far from understanding this system. At present, we know that it is a fairly massive system, consisting of a luminous WR-type star and of a compact object moving in the strong wind emitted by its companion. The interstellar extinction in the direction of Cyg X-3 is very high ($A_{\rm V} \sim 20$), so the optical emission is detectable only in the infrared band. Van Kerkwijk (1992) was the first to investigate the infrared emission from the companion of Cyg X-3 and to find that it looks like a WR-type star. He estimated the spectrum as $\sim$ WN7. The infrared emission was later investigated by Fender et al. (1999) who found that the appearance of the spectrum is variable and is probably modified by the influence of the X-ray emission from the compact object on the stellar wind of the companion. They argued for a rather earlier spectral type: $\sim$ WN4/5. The most recent classification, based on a wide band ($2.4 \div 12 \mu$m) photometric study (Koch-Miramond et al., 2002), favors again a later type ($\sim$ WN8). All these studies suggest that the optical component must be fairly massive ($M_{WR} \ga 5$ M$_\odot$). There was one dissenting opinion: Mitra (1998) proposed a model in which the optical component is a very low mass dwarf. This model is, however, inconsistent with the observed very high infrared luminosity of the object: $M_I \la -4.8$, $M_H \la -5.2$, $M_K \la -5.1$ (Koch-Miramond et al., 2002). At present, there seems to be an agreement that the optical component is, indeed, a luminous WR-type star and that it has to be fairly massive. There is, however, less agreement concerning the mass (and the nature) of the compact object. The orbital phase dependences of the radial velocities of the different infrared emission lines are complicated and open to different interpretations. Assuming one of the possible interpretations, Schmutz et al. (1996) found the mass function $f(M_X) = 2.3 \pm 0.3$ M$_\odot$. Assuming the mass of the W-R star in the range $5 \div 20$ M$_\odot$, this leads to the range $7 \div 40$ M$_\odot$ for the mass of the compact component. According to Schmutz et al., the most likely values are: $M_{\rm WR} \sim 13$ M$_\odot$ and $M_{\rm X} \sim 17$ M$_\odot$. With this interpretation, the compact component must, obviously, be a black hole. On the other hand, Hanson et al. (2000), assuming another interpretation of radial velocities, obtained $f(M_{\rm X}) = 0.027 \pm 0.010$ M$_\odot$ (the value by two orders of magnitude smaller than that of Schmutz et al.!). Hanson et al. conclude that the compact component may be either a black hole with the mass not greater than $\sim 10$ M$_\odot$ or a neutron star.

As may be seen, the status of the compact object in Cyg X-3 cannot be clearly determined at this moment. However, I would like to call attention to the following observational fact. In the case of the mass loss through fast stellar winds (as is the case for W-R stars) the resulting increase of the orbital period must obey the relation $P_{\rm orb}/\dot{P}_{\rm orb} = M_{\rm tot}/\dot{M}_{\rm wind}$, where $M_{\rm tot}$ is the total mass of the binary system (Jeans' mode of mass loss). With the most recent observational determinations: $P_{\rm orb}/\dot{P}_{\rm orb} = (0.95 \pm 0.04) \times 10^6$ yr (Singh et al., 2002) and $\dot{M}_{\rm wind} \sim 1.2 \times 10^{-4}$ M$_\odot$/yr (Koch-Miramond et al., 2002), the implied total mass of the system is $\sim 115$ M$_\odot$. This means, that even taking into account the small precision of the determination of $\dot{M}_{\rm wind}$, the system must consist of two heavy components, probably of a few tens of solar masses each.

To summarize, there is a good chance that Cyg X-3 harbors a black hole (and I would, personally, vote in favor of this option), but the status of the compact component cannot be determined convincingly enough at the present moment.

\section{List of Black Hole Candidates}
\label{sect:lbhc}

Table 2 contains the updated list of the galactic black hole
candidates. The list contains 49 objects. For 19 of them we have
dynamical mass estimate (the last column of the table). These 19
objects may be described as confirmed black holes.

\begin{acknowledgements}
I would like to thank Jorge Casares (acting as a referee) for his
helpful comments. This work was partially supported by the State
Committee for Scientific Research grants No 2 P03C 006 19p01 and
No PBZ KBN 054/P03/2001.
\end{acknowledgements}


\centerline{\bf Tab. 2 $-$ Black Hole Candidates in X-Ray Binaries}
\nobreak
\vspace{5mm}

\vbox{
\begin{tabular}{|rcl|l|l|l|r|r|c|}
\hline
&&&&&&&&\\
\multicolumn{3}{|c|}{Name}&\multicolumn{1}{|c|}{$P_{\rm orb}$}&\multicolumn{1}{|c|}{Opt. Sp} &\multicolumn{1}{|c|}{Other names}&\multicolumn{1}{ |c|}{X$-$R}&\multicolumn{1}{|c|}{C}&\multicolumn{1}{|c|}{$M_{\rm BH}$/ M$_\odot$}\\
&&&&&&&&\\
\hline
&&&&&&&&\\
Cyg X\hspace*{-2.4ex}&$-$&\hspace*{-2.4ex}1&5$^d$6&O9.7 Iab&HDE 226868(O)&pers&$\mu$Q&16 $\pm$ 5\\
&&&&&V1357 Cyg (O)&&&\\
LMC X\hspace*{-2.4ex}&$-$&\hspace*{-2.4ex}3&1$^d$70&B3 V&&pers&&6 $\div$ 9\\
LMC X\hspace*{-2.4ex}&$-$&\hspace*{-2.4ex}1&4$^d$22&O7$-$9 III&&pers&&4 $\div$ 10\\
SS\hspace*{-2.4ex}&&\hspace*{-2.4ex}433&13$^d$1&$\sim$ A7 Ib&V1343 Aql (O)&pers&$\mu$Q&11 $\pm$ 5 ?\\
GX 339\hspace*{-2.4ex}&$-$&\hspace*{-2.4ex}4&1$^d$756&F8$-$G2 III&V821 Ara (O)&RT&&$\ga$ 6\\
3U 0042\hspace*{-2.4ex}&+&\hspace*{-2.4ex}32&11$^d$6&G ?&&T&&\\
XTE J0421\hspace*{-2.4ex}&+&\hspace*{-2.4ex}560&&B[e] I&CI Cam (O)&T&$\mu$Q?&\\
GRO J0422\hspace*{-2.4ex}&+&\hspace*{-2.4ex}32&5$^h$09&M2 V&V518 Per (O)&T&&3.6 $\div$ 5.0\\
&&&&&XRN Per 1992&&&\\
A 0620\hspace*{-2.4ex}&$-$&\hspace*{-2.4ex}00&7$^h$75&K4 V&V616 Mon (O)&RT&&11 $\pm$ 2\\
&&&&&Mon X$-$1&&&\\
&&&&&XN Mon 1975&&&\\
GRS 1009\hspace*{-2.4ex}&$-$&\hspace*{-2.4ex}45&6$^h$96&K8 V&MM Vel (O) &T&&4.4 $\div$ 4.7\\
&&&&&XN Vel 1993&&&\\
XTE J1118\hspace*{-2.4ex}&+&\hspace*{-2.4ex}480&4$^h$1&K7$-$M0 V&KV Uma (O)&T&&$6.0 \div$ 7.7\\
GS 1124\hspace*{-2.4ex}&$-$&\hspace*{-2.4ex}684&10$^h$4&K0$-$5 V&GU Mus (O)&T&&7.0 $\pm$ 0.6\\
&&&&&XN Mus 1991&&&\\
GS 1354\hspace*{-2.4ex}&$-$&\hspace*{-2.4ex}645&&+&BW Cir (O)&T&&\\
&&&&&Cen X$-$2 ?&&&\\
A 1524\hspace*{-2.4ex}&$-$&\hspace*{-2.4ex}617&&+&KZ TrA (O)&RT&&\\
&&&&&TrA X$-$1&&&\\
4U 1543\hspace*{-2.4ex}&$-$&\hspace*{-2.4ex}475&1$^d$12&A2 V&IL Lup (O)&RT&&8.4 $\div$ 10.4\\
XTE J1550\hspace*{-2.4ex}&$-$&\hspace*{-2.4ex}564&1$^d$55&G8 IV$-$K4 III&V381 Nor (O)&RT&$\mu$Q&9.7 $\div$ 11.6\\
4U 1630\hspace*{-2.4ex}&$-$&\hspace*{-2.4ex}472&&+(IR)&Nor X$-$1&RT&&\\
XTE J1650\hspace*{-2.4ex}&$-$&\hspace*{-2.4ex}500&0$^d$212&G$-$K&&T&$\mu$Q &\\
GRO J1655\hspace*{-2.4ex}&$-$&\hspace*{-2.4ex}40&2$^d$62&F3$-$6 IV& V1033 Sco (O)&RT&$\mu$Q&6.3 $\pm$ 0.3\\
&&&&&XN Sco 1994&&&\\
H 1705\hspace*{-2.4ex}&$-$&\hspace*{-2.4ex}250&12$^h$5&K5 V&V2107 Oph (O)&T&&5.7 $\div$ 7.9\\
&&&&&XN Oph 1977&&&\\
XTE J1709\hspace*{-2.4ex}&$-$&\hspace*{-2.4ex}267&&&&T&&\\
GRO J1719\hspace*{-2.4ex}&$-$&\hspace*{-2.4ex}24&&M0$-$5 V&V2293 Oph (O)&T&&\\
&&&&&XN Oph 1993&&&\\
\end{tabular}}

\centerline{\bf Tab. 2 $-$ Black Hole Candidates in X-Ray Binaries (continued)}
\nobreak
\vspace{5mm}

\vbox{
\begin{tabular}{|rcl|l|l|l|r|r|c|}
\hline
&&&&&&&&\\
\multicolumn{3}{|c|}{Name}&\multicolumn{1}{|c|}{$P_{\rm orb}$}&\multicolumn{1}{|c|}{Opt. Sp} &\multicolumn{1}{|c|}{Other names}&\multicolumn{1}{ |c|}{X$-$R}&\multicolumn{1}{|c|}{C}&\multicolumn{1}{|c|}{$M_{\rm BH}$/ M$_\odot$}\\
&&&&&&&&\\
\hline
&&&&&&&&\\
XTE J1720\hspace*{-2.4ex}&$-$&\hspace*{-2.4ex}318&&&&T&&\\
KS 1730\hspace*{-2.4ex}&$-$&\hspace*{-2.4ex}31&&&&T&&\\
GRS 1737\hspace*{-2.4ex}&$-$&\hspace*{-2.4ex}31&&&&T&&\\
GRS 1739\hspace*{-2.4ex}&$-$&\hspace*{-2.4ex}278&&$\ga$ F5 V&&T&&\\
XTE J1739\hspace*{-2.4ex}&$-$&\hspace*{-2.4ex}302&&&&T&&\\
1E 1740.7\hspace*{-2.4ex}&$-$&\hspace*{-2.4ex}2942&12$^d$73&&&pers&$\mu$Q&\\
H 1741\hspace*{-2.4ex}&$-$&\hspace*{-2.4ex}322&&&&T&&\\
SLX 1746\hspace*{-2.4ex}&$-$&\hspace*{-2.4ex}331&&&&T&&\\
XTE J1748\hspace*{-2.4ex}&$-$&\hspace*{-2.4ex}288&&&&T&$\mu$Q&\\
4U 1755\hspace*{-2.4ex}&$-$&\hspace*{-2.4ex}338&4$^h$4 ?&+&V4134 Sgr (O)&pers&&\\
XTE J1755\hspace*{-2.4ex}&$-$&\hspace*{-2.4ex}324&&&&T&&\\
GRS 1758\hspace*{-2.4ex}&$-$&\hspace*{-2.4ex}258&18$^d$45&&&pers&$\mu$Q&\\
SAX J1805.5\hspace*{-2.4ex}&$-$&\hspace*{-2.4ex}2031&&&&T&&\\
XTE J1806\hspace*{-2.4ex}&$-$&\hspace*{-2.4ex}246&&&&T&&\\
XTE J1819\hspace*{-2.4ex}&$-$&\hspace*{-2.4ex}254&2$^d$817&B9 III&V4641 Sgr (O)&T&$\mu$Q&$6.8 \div 7.4$\\
EXO 1846\hspace*{-2.4ex}&$-$&\hspace*{-2.4ex}031&&&&T&&\\
XTE J1856\hspace*{-2.4ex}&+&\hspace*{-2.4ex}053&&&&T&&\\
XTE J1859\hspace*{-2.4ex}&+&\hspace*{-2.4ex}226&9$^h$16&$\sim$ G 5&V404 Vul (O)&T&&$8 \div 10$\\
XTE J1901\hspace*{-2.4ex}&+&\hspace*{-2.4ex}014&&&&RT&&\\
XTE J1908\hspace*{-2.4ex}&+&\hspace*{-2.4ex}094&&+(IR)&&RT&&\\
GRS 1915\hspace*{-2.4ex}&+&\hspace*{-2.4ex}105&33$^d$5&K$-$M III&V1487 Aql (O)&RT&$\mu$Q&14 $\pm$ 4\\
&&&&&XN Aql 1992&&&\\
4U 1918\hspace*{-2.4ex}&+&\hspace*{-2.4ex}146&&&&T&&\\
4U 1957\hspace*{-2.4ex}&+&\hspace*{-2.4ex}115&9$^h$3&+&V1408 Aql (O)&pers&&\\
GS 2000\hspace*{-2.4ex}&+&\hspace*{-2.4ex}251&8$^h$3&K5 V& QZ Vul (O)&T&&7.1 $\div$ 7.8\\
&&&&&XN Vul 1988&&&\\
XTE J2012\hspace*{-2.4ex}&+&\hspace*{-2.4ex}381&&&&T&&\\
GS 2023\hspace*{-2.4ex}&+&\hspace*{-2.4ex}338&6$^d$46&K0 IV& V404 Cyg (O)&RT&&10.0 $\div$ 13.4\\
&&&&&XN Cyg 1989&&&\\
2S 2318\hspace*{-2.4ex}&+&\hspace*{-2.4ex}62&&&&T&$\mu$Q&\\
&&&&&&&&\\
\hline
\end{tabular}}

\vspace{10mm}
\nopagebreak

{\small NOTES:\vspace{2mm}\\
$P_{orb}$ $-$ orbital period\\
Opt. Sp $-$ optical spectrum\\
X-R $-$ X-ray variability\\
C $-$ comments\\
$M_{\rm BH} -$  mass of black hole component\\
$+$  $-$ optical counterpart was identified, but the spectrum was not obtained\\
IR $-$ optical counterpart was seen only in infrared\\
(O) $-$ name of the optical object\\
T $-$ transient\\
RT $-$ recurrent transient\\
pers $-$ persistent\\
$\mu$Q $-$ microquasar\\}

{\small REFERENCES:\vspace{2mm}\\
Most of the references are given in Zi\'o{\l}kowski (2003).
Additional reference is: Hynes et al. (2003) for GX 339$-$4.}

\label{lastpage}

\end{document}